\documentclass[aps,prl,reprint,superscriptaddress,nobibnotes]{revtex4-2}

\usepackage[english]{babel}
\usepackage{xcolor}
\usepackage{comment}
\usepackage{graphicx}
\usepackage{stackengine}
\usepackage{dcolumn}
\usepackage{bm}
\usepackage{bbm}
\usepackage[utf8]{inputenc}
\usepackage{amssymb}
\usepackage{amsmath}
\usepackage{bbold}
\usepackage{pstricks}
\usepackage{slashed}
\usepackage{braket}
\usepackage[hidelinks]{hyperref}
\usepackage{natbib}
\usepackage{float}
\usepackage{verbatim}
\usepackage{siunitx}
\usepackage{lipsum}
\usepackage{slashed}
\usepackage{ulem}
\usepackage[left=1.85cm,right=1.85cm,top=1.85cm,bottom=1.85cm]{geometry}
\usepackage{nicefrac}
\usepackage{cleveref}

\definecolor{mygreen}{rgb}{0,0.5,0}
\definecolor{myblue}{rgb}{0,0,0.75}
\definecolor{mymagenta}{cmyk}{0,1,0,0.12}
\definecolor{mygray}{rgb}{0.5,0.5,0.5}

\renewcommand{\vec}[1]{\boldsymbol{#1}}

\newcommand{\diff}{\mathrm{d}}

\newcommand{\cS}{\mathcal{S}}

\newcommand{\cN}{\mathcal{N}}

\newcommand{\vxy}{\vec{r}_{\perp}}

\newcommand{\vq}{\vec{q}}

\newcommand{\mat}[1]{\mathsf{#1}}

\newcommand{\mS}{\mat{S}}

\begin{document}

\title{Cavity-enhanced continuous-wave microscopy using unstabilized cavities}

\author{Oliver Lueghamer}
\affiliation{Vienna Center for Quantum Science and Technology,
Atominstitut, TU Wien, Stadionallee 2, 1020 Vienna, Austria}

\author{Stefan Nimmrichter}
\affiliation{Naturwissenschaftlich-Technische Fakult{\"a}t, Universit{\"a}t Siegen, Siegen 57068, Germany}

\author{Clara Conrad-Billroth}
\affiliation{University of Vienna, Faculty of Physics, VCQ, A-1090 Vienna, Austria}
\affiliation{University of Vienna, Max Perutz Laboratories,
Department of Structural and Computational Biology, A-1030 Vienna, Austria}

\author{Thomas Juffmann}
\affiliation{University of Vienna, Faculty of Physics, VCQ, A-1090 Vienna, Austria}
\affiliation{University of Vienna, Max Perutz Laboratories,
Department of Structural and Computational Biology, A-1030 Vienna, Austria}

\author{Maximilian Pr\"{u}fer}
\email[Corresponding author: ]{maximilian.pruefer@tuwien.ac.at}
\affiliation{Vienna Center for Quantum Science and Technology,
Atominstitut, TU Wien, Stadionallee 2, 1020 Vienna, Austria}

\begin{abstract}
Microscopy gives access to spatially resolved dynamics in different systems, from biological cells to cold atoms. A big challenge is maximizing the information per used probe particle to limit the damage to the probed system. We present a cavity-enhanced continuous-wave microscopy approach that provides enhanced signal-to-noise ratios at fixed damage. Employing a self-imaging 4f cavity, we show contrast enhancement for controlled test samples as well as biological samples. For thick samples, the imaging cavity leads to a new form of dark-field microscopy, where the separation of scattered and unscattered light is based on optical path length. We theoretically show that enhanced signal, signal-to-noise, and signal-to-noise per damage are also retrieved when the cavity cannot be stabilized. Our results provide an approach to cavity-enhanced microscopy with unstabilized cavities and might be used to enhance the performance of dispersive imaging of ultracold atoms.

\end{abstract}

\maketitle
Advances in microscopy have led to new discoveries across scientific disciplines \cite{hawkes2007science}, from cellular biology \cite{morris2019microscopy} to quantum physics \cite{ketterle_making_1999} and beyond. Microscopy setups measure certain parameters of interest in a spatially resolved way. Not only the spatial resolution limits the measurement precision but often the limiting factor is the finite number of probe particles detected in a given image. In many applications, the number of probe particles cannot be increased arbitrarily, either due to sample, source, or detector restrictions. Maximizing the information obtained from each detected probe particle is crucial in such applications. 

For coherent imaging techniques, it has recently been shown that multi-passing each probe particle through the sample can increase the information per detected probe particle. This has been experimentally demonstrated in optical imaging and diffraction studies~\cite{juffmann_multi-pass_2016}, where self-imaging cavities~\cite{arnaud_degenerate_1969} were used to multi-pass a pulsed probe through a sample. Follow-up experiments showed the build-up of orbital angular momentum in a multi-passing experiment~\cite{Klopfer2016}. Notably, multi-passing probe particles $m$ times through a sample can enable a measurement precision per probe-sample interaction similar to a quantum-enhanced measurement with $m$ suitably entangled probe particles~\cite{luis_phase-shift_2002, Giovannetti2006a, Higgins2007a, Koppell2022}.

In physics applications, like spectroscopy or the imaging of ultracold atoms, a narrow spectral linewidth is often required. Thus, cavity-enhanced microscopy has to be done with continuous-wave (CW) excitation, which actually has theoretically been shown to outperform the pulsed multi-pass scheme \cite{nimmrichter_full-field_2018}. However, despite the widespread use of single-mode cavity-enhanced measurements in various scientific and technological fields, applications in imaging remain rare, due to challenges in operating a cavity that is fully degenerate in all transverse modes~\cite{gigan_image_2005}. The first progress was recently demonstrated in a 4-pass geometry that showed contrast enhancement in flow-cytommetry~\cite{israel_continuous_2023}. So far, genuine continuous-wave imaging cavities have been employed in the realization of multi-mode lasers~\cite{Cao2019}, multi-mode coherent absorbers~\cite{slobodkin_massively_2022}, and cavity-enhanced non-linear optics applications~\cite{Kolobov2007a}. 

In this paper, we demonstrate cavity-enhanced continuous-wave microscopy using a self-imaging $4f$ cavity. 
We demonstrate the concept experimentally by applying it to the imaging of an artificially fabricated sample, consisting of holes in a $10\,\text{nm}$ thin silicon nitride ($\text{Si}_3\text{N}_4$) membrane. Imaging human cheek cells, we further demonstrate that cavity-based imaging leads to a novel form of dark-field microscopy, where the separation of scattered and unscattered light is based on optical path length and not on scattering angle. Our experiments are backed by theory which shows that an enhanced contrast can even be observed when scanning the cavity across a free spectral range of the cavity. These results open a path to cavity-enhanced measurements in applications where cavity stabilization is impossible. Our continuous-wave approach will facilitate the usage of cavities for microscopy with the need for small spectral width, for example, the imaging of ultracold atoms. 

\begin{figure*}[t]
	\centering
	\includegraphics[width=\linewidth]{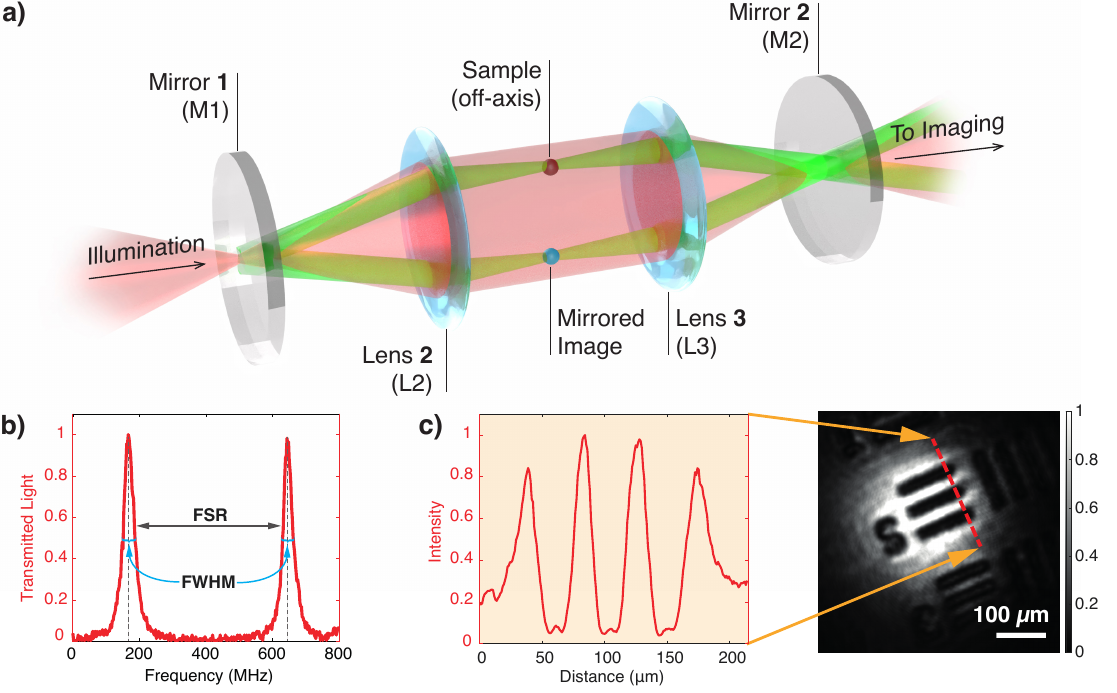}   
	\caption{ \textbf{Cavity setup and characterization. a)} Ray tracing picture showing the self-imaging properties of the 4f cavity. The sample is illuminated with a collimated beam (red); the diffracted light (green) is imaged onto the sample after one full roundtrip. \textbf{b) } Cavity length scan monitored by the output voltage (normalized) of a photodiode placed at the output of the cavity. The separation between the two resonance peaks yields a free spectral range (FSR) of $488\, \text{MHz}$. \textbf{c) } We test the degeneracy of transversal modes by projecting the image of a US Air Force (USAF) Target onto the cavity image plane. The recorded light intensity in a conjugate plane after the cavity, and a cross-section thereof, demonstrate its image transmission capabilities.}
	\label{fig:Cavity}
\end{figure*}

\begin{figure*}[t]
	\centering
	\includegraphics[width=\linewidth]{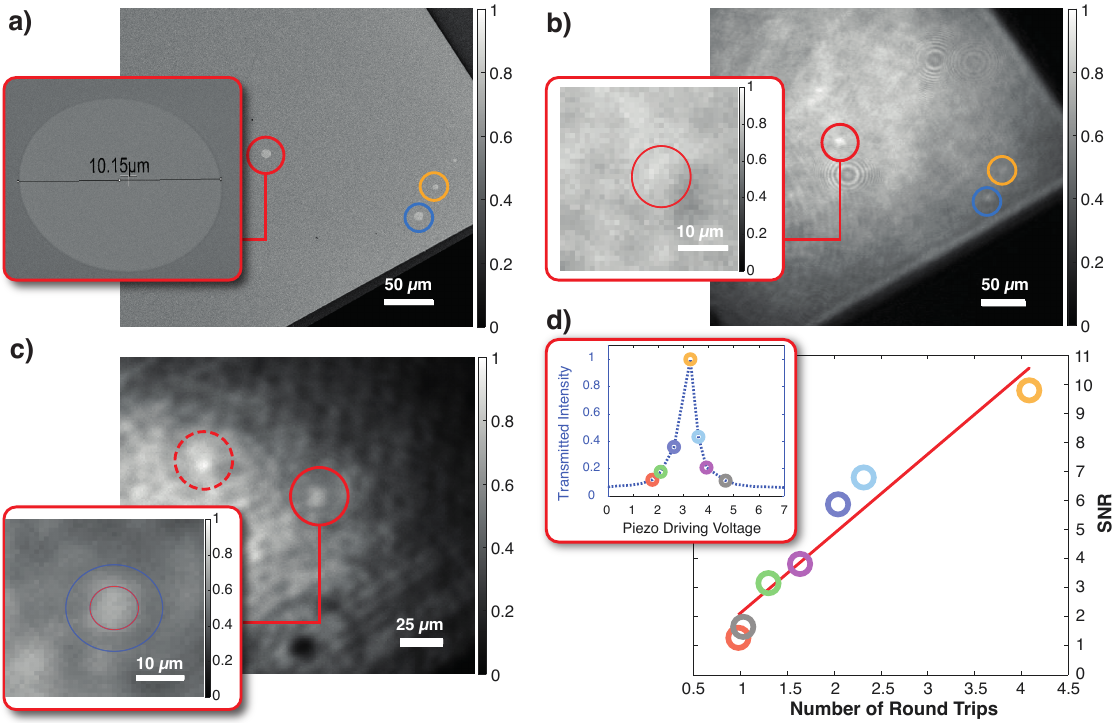} 
	\caption{\textbf{Cavity-enhanced microscopy of holes in a $\text{Si}_3\text{N}_4$ membrane}. \textbf{a)} Scanning electron microscopy image of holes cut into a $\text{Si}_3\text{N}_4$ membrane. The red and blue circles indicate holes with a diameter of $10\,\mu\text{m}$, and the yellow circle marks a hole with a diameter of $6\,\mu\text{m}$. The inset shows a close-up of the hole we use as our main target. \textbf{b)} Defocused single-pass image of the membrane; the inset shows the region of the image with the target in focus; the target hole is not visible (see main text). \textbf{c)} Cavity-enhanced image of the membrane; the two circles indicate the targeted hole (solid) and its reflection (dashed line). The inset shows a close-up of the target; compared to the single pass case we find a contrast enhancement; the two circles show the regions for averaging signal (red) and background (blue without red) intensities, used to estimate contrast and SNR. \textbf{d)} Signal-to-noise ratio in the cavity-enhanced case plotted against the number of effective round trips. The color-coded data points match those in the inset, where the transmitted intensity is shown as a function of the cavity length.}
	\label{fig:Holes}
\end{figure*}

\textit{Degenerate cavity setup.---}
For cavity-enhanced microscopy, it is crucial that spatial information is accurately transmitted through the cavity system. 
To achieve this, the transverse cavity modes have to be degenerate, i.e. they resonate simultaneously at the same cavity length. Degeneracy can be achieved by using lenses within the cavity to ensure that any arbitrary ray retraces its path after a full round trip~\cite{arnaud_degenerate_1969}. 
We employ an effective 4f configuration using two highly reflective cavity mirrors and a pair of $f = 75\,\text{mm}$ biconvex lenses; a schematic of the setup is shown in \crefformat{figure}{Fig.~#2#1{a}#3}\cref{fig:Cavity} (see Supplementary Material for details). The sample is positioned off-axis to prevent overlapping with the mirrored image; as a result, the image is projected onto the sample only after completing one full round trip. Control over the cavity length is achieved via a piezo ring mounted on the first mirror. For monitoring the cavity output, a photodiode is placed after a beamsplitter, and the remaining light is imaged on a charge-coupled device (CCD) camera. Various standard imaging modalities, such as bright-field, darkfield, or Zernicke phase imaging, can be implemented outside the cavity (see Supplementary Material for imaging details).

In \crefformat{figure}{Fig.~#2#1{b}#3}\cref{fig:Cavity}, we present a cavity scan where we show the output intensity measured using a photodiode as a function of the cavity length. The spacing between the two peaks represents the free spectral range (FSR), achieved by moving the mirror over half a wavelength $\lambda$. A frequency ruler is implemented via a two-frequency method. From the distance between the two peaks, we obtain an FSR of $488\, \text{MHz}$, which aligns well with the theoretical predictions (see Supplementary Material for details).

To demonstrate image transmission capabilities, a US Air Force (USAF) target is positioned in the object plane of the focusing lens {in front of the cavity}. The resulting image is coupled into the self-imaging cavity and then captured by the CCD camera. As shown in \crefformat{figure}{Fig.~#2#1{c}#3}\cref{fig:Cavity}, the image clearly reveals the structure of the USAF target, confirming that many transverse modes are transmitted through the cavity. 

\textit{Theoretical results for cavity-enhancement---}
The theoretical performance of our approach is analyzed in a {calculation} that keeps track of phase shifts along each closed path within the cavity (see Supplementary Material). 
Our key performance indicators are the sample contrast and the signal-to-noise ratio (SNR) at fixed input power,
\begin{equation}
\mathcal{C} = \frac{|I_S - I_B|}{I_S + I_B}, \qquad 
    \text{SNR} = \frac{|I_{S}-I_{B}|}{\sqrt{I_{S}+I_{B}}},
    \label{C_and_SNR}
\end{equation}
where $I_{S}$ is the output intensity measured at the sample position, and $I_{B}$ is the average output intensity measured next to the sample.

While in phase-contrast microscopy the contrast would be $\mathcal{C}_{PC} = 2|(\chi-\chi_0)|$~\cite{bouchet2021}, we find for our cavity-enhanced approach:
\begin{align}
    \mathcal{C}_{\max} &\approx \frac{2\sqrt{R_2}}{T_1+T_2} |(\chi-\chi_0)\sin 4kf|, 
    \end{align}
where $R_i$ and $T_i$ denote the reflectivity and transmission of the cavity mirrors $i\in {1,2}$, $\chi$ and $\chi_0$ are the weak phase shifts of the sample and a reference point, respectively, and $k=2\pi/\lambda$. Note that the prefactor is $>>1$ if high-reflectivity mirrors are used ($T_i<<1$).

Strikingly, when averaging over the resonance in the course of the measurement, the contrast is still enhanced compared to a single pass:
\begin{align}
    \mathcal{C}_{\rm avg} &\approx  \frac{\sqrt{R_2}}{T_1+T_2} |(\chi-\chi_0)\sin 4kf| = \frac{\mathcal{C}_{\max}}{2} ,
\end{align}
Note that in the shot-noise limited case the signal-to-noise ratio at constant damage SNR$_D$ will also be enhanced \cite{nimmrichter_full-field_2018}.

This enhanced performance of an unstabilized cavity relies on all modes within the cavity undergoing the same fluctuations. It can also be derived from a textbook-like toy model in which a Fabry-Pérot cavity is used for gas cell spectroscopy. If the input mirror vibrates across an FSR, cavity enhancement is still obtained, if the cavity output is interfered with a reference from an empty cavity that undergoes the same fluctuations in cavity length. We provide this calculation in the Appendix to this manuscript, where we also show that the 2x loss in contrast for unstabilized cavities is due to light being reflected from the cavity. Analyzing the reflected light, will regain the lost information. 

\begin{figure}[t]
	\centering
	\includegraphics[width=\linewidth]{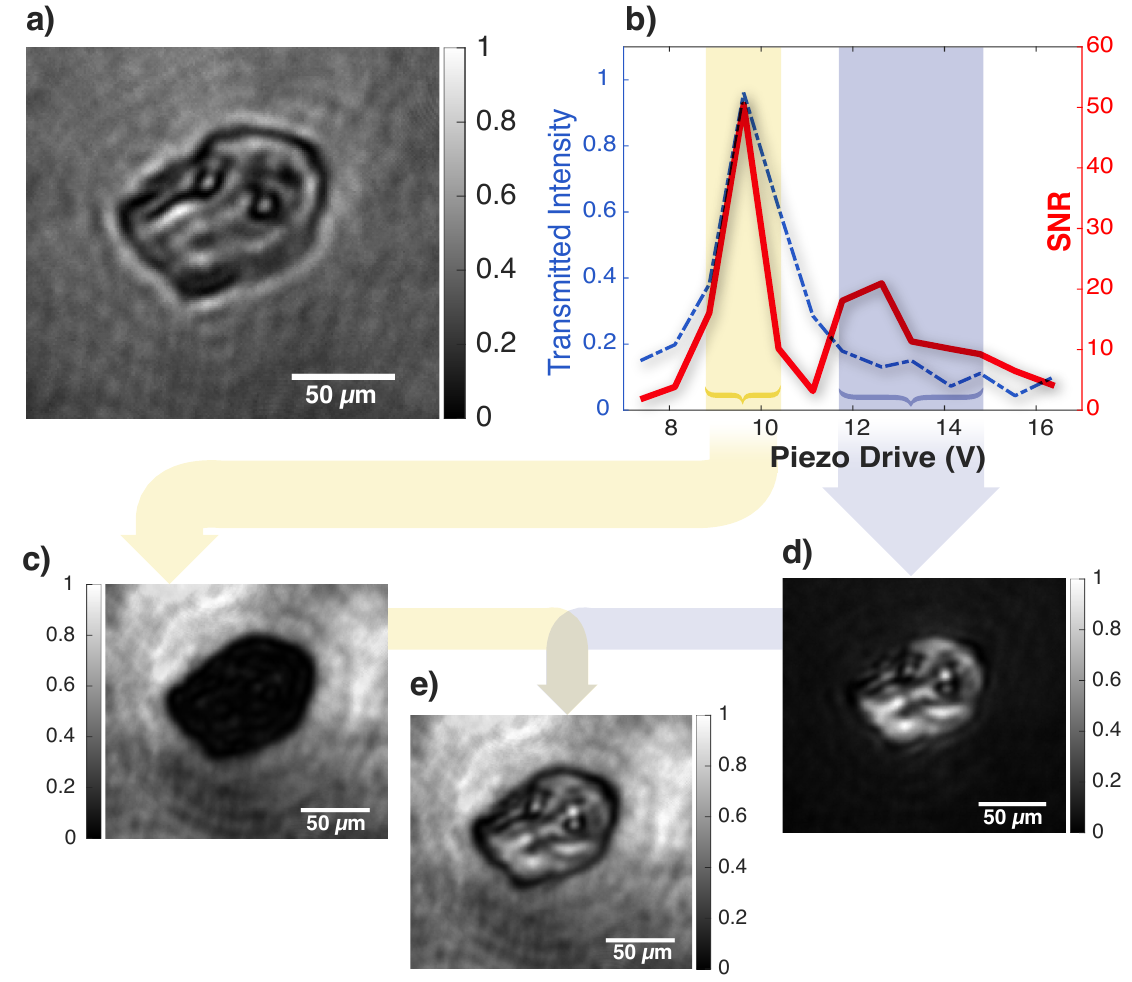} 
	\caption{ \textbf{Cavity dark-field imaging. a)} Bright Field image of cheek cell in single pass. \textbf{b)} Calculated SNR as a function of mirror displacement represented by the driving voltage of the piezo. The blue curve shows the cavity output detected by the photodiode. \textbf{c)} Yellow: Integrated images over time, while the incidence light is resonant and the refracted light is out of resonance. \textbf{d)} Dark Blue: Integrated images over time while incidence light is out of resonance and refracted light on resonance. This represents a novel approach to dark-field imaging. \textbf{e)} Integrating over all the detunings, we obtain a picture corresponding to an unstabilized cavity with increased contrast as compared to the single-pass image.}
	\label{fig:Cheek}
\end{figure}

\textit{Demonstration of cavity-enhanced microscopy.---} The enhancement of multi-pass techniques is closely linked to the effective number of round trips \textit{N}~\cite{juffmann_multi-pass_2016, nimmrichter_full-field_2018, israel_continuous_2023}.
For continuous-wave cavities, we can calculate {the effective} \textit{N} by comparing the intensities inside and outside the cavity for a given cavity length,
\begin{equation}
    N = \frac{\sqrt{I_{\rm cav}}}{\sqrt{T_1 I_{\rm in}}} = \frac{\sqrt{I_{\rm out}}}{\sqrt{T_1 T_2 I_{\rm in}}}\,.
    \label{N_int}
\end{equation}
Measuring the total input and output intensities $I_{\rm in}, I_{\rm out}$ thus allows us to evaluate $N$ for a given cavity setting in the experiment.

In order to demonstrate the contrast-enhancing capabilities of the setup, we cut holes with diameters of $6\ \&\ 10\,\mu\text{m}$ in a $10\,\text{nm}$ thin silicon nitride ($\text{Si}_3\text{N}_4$) membrane, which forms a $500\times500\,\mu\text{m}$ window on a silicon frame; the membrane has a reflectance of approximately $1.5\,\%$.
\crefformat{figure}{Fig.~#2#1{a}#3}\cref{fig:Holes} shows a scanning electron microscopy image of the structure. 
In \crefformat{figure}{Fig.~#2#1{b}#3}\cref{fig:Holes} we show a single-pass defocused image~\cite{defocusreview}, where the hole is visible as a bright spot. The inset shows a focused single-pass microscopy image in which the holes are not visible. This is consistent with the contrast levels expected for these techniques. While bright field-contrast would be based on reflection from the membrane and amounts to only $1.5\%$, defocus leads to phase contrast which can be as large as $\Delta I/I=2(\chi-\chi_0)=(n_{\rm SiN}^2-1)kd=24\%$, where $n_{\rm SiN}$ denotes the index of refraction of $\text{Si}_3\text{N}_4$; see supplementary equation \eqref{eq:sampleParam}.

We find a noticeable contrast enhancement in the cavity-enhanced image on resonance (\crefformat{figure}{Fig.~#2#1{c}#3}\cref{fig:Holes}). To quantify the contrast, we averaged the intensity of the cavity-enhanced hole and compared it to the nearest averaged radial background.
The calculated contrast was approximately $10\%$, considerably more than the single-pass bright-field contrast. 
We then evaluated the SNR for each captured image, plotting it against the effective number of roundtrips. Remarkably, we observed an enhanced SNR at the cavity resonance, as depicted in \crefformat{figure}{Fig.~#2#1{d}#3}\cref{fig:Holes}, with the overall SNR behavior following the cavity output (see inset of \crefformat{figure}{Fig.~#2#1{d}#3}\cref{fig:Holes}). Further characterization of the image reveals a measured hole size of $\approx 11\,\mu m$ and a resolution of $3.05\mu m$, which aligns well with the expected values (see Supplementary Material), indicating that the cavity does not reduce the spatial resolution of our microscope.

\textit{Microscopy of epithelial cells.---}
Next, we image epithelial cells obtained from the inner lining of a human cheek, which are fixed onto an anti-reflection coated glass slide using methanol. \crefformat{figure}{Fig.~#2#1{a}#3}\cref{fig:Cheek} displays the morphology of these cells in a single-pass image, with the two black circles in the center likely representing the cell nuclei.

Performing cavity-enhanced microscopy, we first study the SNR as a function of cavity length, defining the whole cell as the signal region and the region around it as the background. We obtain a peak in SNR when the incident, unscattered background light is resonant (indicated by the yellow bar in Fig.\,\ref{fig:Cheek}). Notably, the transparent sample appears completely darkened (see \crefformat{figure}{Fig.~#2#1{c}#3}\cref{fig:Cheek}). Detuning the cavity, we observe a second “resonance” in the SNR when the cavity is off-resonant for the incident light. In \crefformat{figure}{Fig.~#2#1{d}#3}\cref{fig:Cheek}, we show the resulting image, which clearly displays the cell content and its structures spatially resolved (see Supplementary Material for all images). The phase shift that the light undergoes while passing through the sample, along with the laser's narrow bandwidth, is substantial enough for the refracted light to resonate at a different cavity length. This phenomenon introduces a new method for dark-field imaging; the peak shift can be used as a measure of the sample's optical thickness.

\textit{Microscopy with unstabilized cavities.---}
Finally, we show that signal amplification, and enhanced contrast, are also observed in an unstabilized cavity. 
To demonstrate this effect experimentally, we integrate across the entire measurement, and obtain a combination of bright- and darkfield image, as shown in \crefformat{figure}{Fig.~#2#1{e}#3}\cref{fig:Cheek} with enhanced contrast compared to single-pass. This effect is also achieved by integrating with the exposure time over one FSR scan.
This makes our cavity-enhanced microscopy method applicable in scenarios without the possibility of stabilizing the cavity actively.

\textit{Conclusion and outlook.---}
In this work, we demonstrate continuous-wave cavity-enhanced microscopy with a narrow laser bandwidth. We experimentally demonstrate increased contrast and signal-to-noise, and show both theoretically and experimentally that these advantages persist in unstabilized cavities. 

Furthermore, we present a new form of dark-field microscopy in which different regions of a sample, corresponding to different optical path lengths, light up at different cavity lengths. This corresponds to an entirely new contrast mechanism in optical microscopy. Contrary to traditional dark-field imaging, our scheme enables dark-field imaging with forward scattered light, which is ideal for quantitatively assessing samples characterized by slowly varying phase shifts. 

A potential application of our method is cavity-enhanced imaging of ultracold atoms, which so far has only been done with single mode cavities \cite{mazzinghi_cavity-enhanced_2021}. Dispersive imaging, utilizing the phase shifts \cite{ketterle_making_1999}, has been demonstrated to allow for multiple repeated images \cite{andrews_direct_1996}. We propose using cavity-enhanced imaging to improve the signal-to-noise ratio at fixed feedback to the quantum system to lower the impact of e.g. heating \cite{altuntacs2023quantum}. The forward scattering leading to the phase signal is coherently amplified by the cavity, in contrast to incoherent scattering processes. This should enable high SNR dispersive imaging of ultracold atoms. At the same time, the imaging cavity provides new ways of applying local control fields within the cavity \cite{KroezeCavityMicro2023}.

\begin{acknowledgements}

\section{Acknowledgments} We thank Helmut Hörner, Julian Léonard, Stefan Rotter, and Jörg Schmiedmayer for discussions. M.P. thanks Jörg Schmiedmayer for scientific guidance and for providing the laboratory infrastructure. The $\text{Si}_3\text{N}_4$ sample preparation was carried out using facilities
at the University Service Centre for Transmission Electron Microscopy, Vienna University of Technology, Austria. This work is supported by an ESQ
Discovery grant by the Austrian Academy of Sciences (QUIMP; 10.55776/ESP396). M.P. has received funding from Austrian Science Fund (FWF): ESP~396 (QuOntM). T.J. and C.B. acknowledge funding from the European Union's Horizon 2020 research and innovation programme under grant agreement 758752. 

~

~
M.P. and T.J. conceived the measurement scheme. C.B. took preliminary data. O.L. built the experimental setup and took the experimental data with the help of M.P..  O.L., T.J., and M.P. analyzed the experimental data. S.N. and T.J. developed the theoretical model. All authors contributed to the discussion of the results. O.L., S.N., T.J., and M.P. wrote the manuscript.

\end{acknowledgements}

\newpage
	\appendix
	\section*{Appendix}

\textit{Cavity parameters.---}
The cavity length is defined by the $4f$ requirement to be $\sim 30\,\text{cm}$. Accounting for the optical path length differences introduced by the lenses results in an actual length of approximately $30.6\,\text{cm}$. This corresponds to an FSR of $493\,\text{MHz}$, which is in good accordance with the experimentally found value of $\approx 488\,\text{MHz}$. We use two highly reflective mirrors with $R_1 = 95\,\%$ and $R_2 = 86\,\%$, yielding a theoretical finesse value of $29.8$. 

In the actual experimental environment, we observed fluctuating finesse values between $10$ and $25$. The presence of optical elements within the cavity and the sample carriers likely contributed to this reduction in overall finesse. We observed cavity length drifts of about $\pm 30\,\text{MHz}$ on the one second scale.

\textit{Enhancement with unstable cavity.---}
In the main text, we demonstrate cavity-enhanced microscopy of an optically thin lossless sample and show evidence that an enhancement of the phase contrast prevails even in the case of an unstable cavity resonance. 
Here, we corroborate our finding with the help of an instructive toy model based on a Fabry-P\'{e}rot cavity - a setup known from textbooks and many applications. This simple example illustrates both the contrast enhancement and its robustness against cavity fluctuations without lengthy formulas.

\begin{figure}[t]
	\centering
	\includegraphics[width=\linewidth]{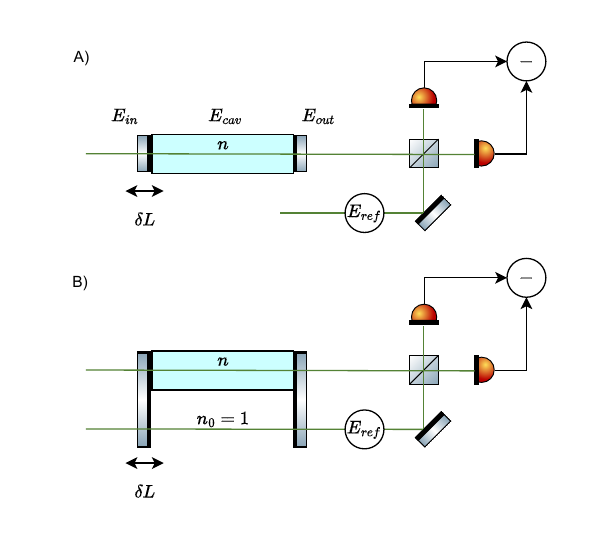}  
	\caption{\textbf{Set-up for toy model calculation}. \textbf{a) }The output of a Fabry-Pérot cavity setup is homodyned with a local oscillator. \textbf{b) } Additionally, the local oscillator is transmitted through the same unstable cavity picking up the phase shift of the detuned cavity. For both cases, we analyze the possible precision in estimating the phase shift imprinted by a thin lossless sample inside the cavity.}
	\label{fig:EnhancementUnstable}
\end{figure}

Consider a Fabry-P\'{e}rot cavity of length $L$ with end mirrors of reflectivity $R = 1-T$ that contains a homogeneous, lossless, and optically thin medium. The task is to infer its weak refractive index $n$ from a measurement of the small phase shift, $\chi = (n-1)k L \ll 1$, it imparts on a {single-mode} light field passing through the cavity. A balanced beam splitter then superimposes a {phase-locked} reference beam on this field, after which the intensities in the two output ports are detected; see Fig.~\ref{fig:EnhancementUnstable}. We will take their difference as the measurement signal. 

Let us assume for now that the cavity length is stable. Given a monochromatic input field amplitude $E_{\rm in}$ of wavenumber $k = 2\pi/\lambda$, the transmitted field reads as \cite{Hodgson2005}
\begin{equation}\label{eq:Eout}
    E_{\rm out} = \frac{T E_{\rm in} e^{ik(n-1)L}}{1 - R e^{2iknL} } \approx \sqrt{T}E_{\rm cav} \left[ 1 + i\chi \frac{1+Re^{2ikL}}{1-Re^{2ikL}} \right].
\end{equation}
Here and in the following, we consistently expand to lowest order in the phase shift $\chi$ we seek to estimate, and we denote the field amplitude inside the \textit{empty} cavity as
\begin{equation}\label{eq:Ecav}
    E_{\rm cav} = \frac{\sqrt{T} E_{\rm in}}{1-R e^{2ikL}}.
\end{equation}
Notice that all field amplitudes are defined with respect to the position of the right cavity mirror on the optical axis, $z\equiv 0$.

We achieve perfect, resonant transmission if we adjust the cavity length or the light wavelength such that the optical path length $2L$ over one roundtrip is a multiple of the wavelength, $kL = \ell \pi$ with $\ell \in \mathbb{N}$. This leads to constructive interference enhancing both the intra-cavity field strength, $E_{\rm cav}^{\rm res} = E_{\rm in}/\sqrt{T} > E_{\rm in}$, and the response to the phase shift in \eqref{eq:Eout},
\begin{equation}
    E_{\rm out}^{\rm res} = \sqrt{T} E_{\rm cav}^{\rm res} \left[ 1 + i\chi \frac{1+R}{T}\right].
\end{equation}
For phase-sensitive detection, we combine the output with a reference beam of, say, the same input field strength phase-shifted by $\pi/2$, $E_{\rm ref} = iE_{\rm in}$, as depicted in Fig.~\ref{fig:EnhancementUnstable}(a).  The intensities detected after the beam splitter are given by $|E_{\rm ref} \pm E_{\rm out}|^2/4$, up to constant prefactors, and the measurement signal by the difference of these two intensities,
\begin{align}
    \cS_{\rm a} &\propto \mathrm{Re} \left\{ E_{\rm ref}^* E_{\rm out} \right\} \nonumber \\
    &\approx T|E_{\rm in}|^2 \mathrm{Im} \left\{ \frac{1+i\chi \frac{1+Re^{2ikL}}{1-Re^{2ikL}}}{1-Re^{2ikL}} \right\}. \label{eq:S}
\end{align}
The resonant case yields a gain in phase signal due to the enhancement of the intra-cavity intensity with $1/T$,
\begin{equation}
    \cS^{\rm res} \propto |E_{\rm in}|^2 \frac{1+R}{T}\chi = T|E_{\rm cav}^{\rm res}|^2 \frac{1+R}{T}\chi. \label{eq:Sres}
\end{equation}
Not only the signal is enhanced on resonance, but also the signal-to-noise ratio (SNR): assuming shot-noise-limited detection, the standard deviation $\cN$ of the measurement signal is proportional to the square-root of the sum of the two detected intensities. 
In the resonant case, $\cN^{\rm res} \propto |E_{\rm in}|$, and hence the SNR improves with $1/T$. 

For sensitive specimens, it is important to characterize the SNR at a fixed intracavity intensity, i.e., a fixed amount of probe-induced specimen damage. To do so, we compare the cavity-enhanced case with a single pass scheme with an input power increased by $(1+R)/T$. We retrieve a signal-to-noise ratio at constant damage (SNR$_D$) that improves by $1/\sqrt{T}$ \cite{nimmrichter_full-field_2018}, indicating another fundamental advantage of cavity-enhanced measurements over single-pass measurements.

What happens if the cavity length $L$ is unstable for the duration of the experiment? We can describe this by taking uniform averages with respect to $L$ over one spectral range $\lambda/2$, neglecting the small variation of the phase shift $\chi$ over this range. Analytic expressions for the averaged intracavity intensity and the measurement signal are readily obtained with help of Mathematica,
\begin{align}
    \overline{|E_{\rm cav}|^2} &= \frac{|E_{\rm in}|^2}{1+R}, \\
    \overline{\cS}_{\rm a} &\propto T|E_{\rm in}|^2 \chi = T \overline{|E_{\rm cav}|^2} (1+R) \chi . \label{eq:Savg_bad}
\end{align}
The average intracavity intensity is no longer enhanced, but \textit{lower} than the input intensity, reducing the signal proportionally. However, even if we increase the input power to match the intra-cavity intensity of the resonant signal \eqref{eq:Sres}, the sensitivity to the phase shift is still diminished, lacking the resonant gain by $1/T$. In agreement with our intuition, we do not observe an enhanced signal, SNR, and SNR$_D$ with a cavity of unstable length.

However, the situation changes if we let the reference beam pass an empty region of the same cavity and thus be subject to the same unstable length as the probe field; see Fig.~\ref{fig:EnhancementUnstable}(b). This is described by the amplitude $E_{\rm ref} = i\sqrt{T}E_{\rm cav}$, which leads to the measurement signal 
\begin{equation}
    \cS_{\rm b} \propto T|E_{\rm cav}|^2 \mathrm{Re} \left\{ \chi \frac{1+Re^{2ikL}}{1-Re^{2ikL}}\right\} \label{eq:S2}
\end{equation}
On resonance ($kL = \ell \pi$), the result is the same as before in \eqref{eq:Sres}. The length-averaged signal for an unstable cavity now reads as
\begin{equation}
    \overline{\cS}_{\rm b} \propto |E_{\rm in}|^2 \frac{1+R^2}{(1+R)^2}\chi = T\overline{|E_{\rm cav}|^2} \frac{1+R^2}{(1+R)T}\chi. 
\end{equation}
Once again, the signal suffers from the reduced intra-cavity intensity, but at matching intensity, the gain in phase sensitivity persists. At high cavity finesse, $R\approx 1$, the sensitivity is reduced merely by a factor two compared to the resonant case \eqref{eq:Sres}. Regarding SNR, the detected intensities are diminished by $T$ compared to the input, so that $\mathcal N \propto \sqrt{T}$ and hence SNR $\propto 1/\sqrt{T}$. Compared to a single pass, the intracavity intensity is now reduced by $1/(1+R) \approx 1/2$, while the detected intensity is reduced further by $T$. Hence, at matching damage, SNR$_D$ improves by $\sqrt{2/T}$.

In the Supplementary Material, we analyze an imaging scenario in which many transverse modes are enhanced in a self-imaging cavity. Since they all are transmitted through the same cavity, the assumption of having common noise on the cavity path length is justified for small path length changes on the order of $\lambda$ and in the paraxial limit. Spatially distributed phase shifts can be quantified using phase contrast microscopy, which can be analyzed as an interference effect between the scattered and unscattered light. The imaging setup thus plays a role similar to the homodyne detection in our toy model. Also there, we find enhanced signal contrast, which implies, by virtue of the same arguments as made here, that the SNR and SNR$_D$ are enhanced compared to single-pass imaging as well.

\textit{Information in reflected field.---}
We have seen that, in the limit of high cavity finesse, the length average leads to a loss in sensitivity by two in case (b) compared to the resonant result. The intuition is that this should come from the signal that is transmitted through the left cavity mirror. The total outgoing field from the left mirror to the left is
\begin{equation}
    E_{\rm out}^{\leftarrow} = \sqrt{R} E_{\rm in} \left[1-\frac{Te^{2inkL}}{1-Re^{2inkL}} \right] = \frac{\sqrt{R} E_{\rm in} (1-e^{2inkL})}{1-Re^{2inkL}},
\end{equation}
and one can easily check the conservation of total energy flux, $|E_{\rm in}|^2 = |E_{\rm out}|^2+|E_{\rm out}^{\leftarrow}|^2$. For weakly refractive media, we again expand to lowest order in $\chi$,
\begin{equation}
    E_{\rm out}^{\leftarrow} \approx \sqrt{R}E_{\rm in} \left[ \frac{1-e^{2ikL}}{1-Re^{2ikL}} - 2i\chi \frac{Te^{2ikL}}{(1-Re^{2ikL})^2} \right].
\end{equation}
Phase-sensitive detection of this field with help of an empty reference cavity of the same varying length is achieved in the same manner as for the outgoing field on the right. Given the phase-shifted reference field $E_{\rm ref}^{\leftarrow} = iE_{\rm out}^{\leftarrow}|_{\chi=0}$ of the empty cavity, we have
\begin{align}
    \mathcal{S}_{\rm b}^{\leftarrow} &\propto \mathrm{Re} \left\{ E_{\rm ref}^{\leftarrow*} E_{\rm out}^{\leftarrow} \right\} \nonumber \\
    &= \frac{2T R (1+R) |E_{\rm in}|^2}{|1-Re^{2ikL}|^4}(1-\cos 2kL)\chi .
\end{align}
Averaging this over the length variation results in
\begin{equation}
    \overline{\mathcal{S}_{\rm b}^{\leftarrow}} \propto |E_{\rm in}|^2 \frac{2R}{(1+R)^2}\chi .
\end{equation}
Adding up both average signals, we get $\overline{\mathcal{S}}_{\rm b} + \overline{\mathcal{S}_{\rm b}^{\leftarrow}} \propto |E_{\rm in}|^2 \chi = \overline{|E_{\rm cav}|^2} (1+R) \chi$. Hence, if we increase the input intensity to match the $1/T$-enhanced intracavity intensity of the resonant case \eqref{eq:Sres}, then for $R\approx 1$ the total sensitivity to $\chi$ is the same. Indeed, the missing factor of two in the forward signal is recovered by also monitoring the output on the left.

\newpage
~
\newpage
\onecolumngrid
\appendix
\section*{Supplementary Material}
\setcounter{figure}{0}
\renewcommand{\figurename}{\textbf{Supplementary Material Fig.}}

\section{Experimental setup}
\begin{figure*}[t]
	\centering
	\includegraphics[width=\linewidth]{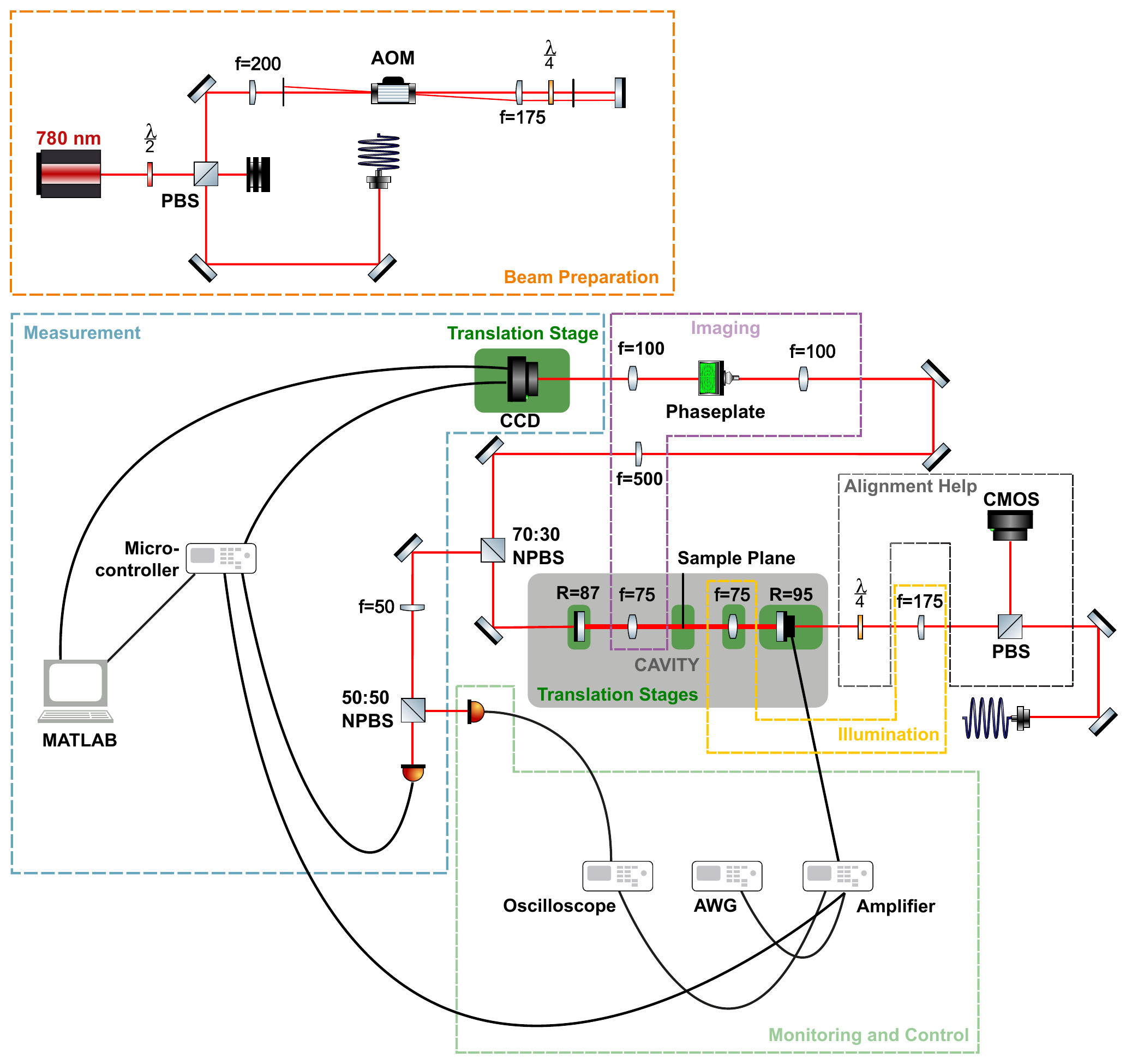}  
	\caption{\textbf{Experimental Setup} A laser at $780\, $nm is coupled into a single-mode fiber following beam preparation, including power regulation and a double-pass acousto-optic modulator (AOM) setup for measuring the free spectral range (FSR) and potentially pulsing the laser. Subsequently, it encounters an alignment help, and a first lens focuses the beam on the in-coupling mirror, allowing for full-field illumination at the sample plane. Upon entering the cavity, the output is detected by two photodiodes and imaged by a camera after magnification, offering the possibility of implementing dark-field and phase-contrast imaging via a 4f setup. For real-time monitoring of the cavity response, the second photodiode is connected to an oscilloscope. To scan the cavity, the in-coupling mirror is affixed to a piezoelectric ring, enabling effective adjustment of the resonator's length.The focal lengths $f$ are given in mm, the reflectivities $R$ are given in \%.}
	\label{fig:setup}
\end{figure*}

In \crefformat{figure}{Fig.~#2{A}#1#3}\cref{fig:setup}, we show the complete setup, consisting of three main parts. First, the $780\, $nm laser undergoes beam preparation, depicted at the top. Power regulation without altering the laser current is achieved using a half-wave plate ($\nicefrac{\lambda}{2}$) and a polarizing beamsplitter (PBS). A double-pass acousto-optic modulator (AOM) adjusts frequency, with ruler frequencies enabling free spectral range (FSR) measurement.

After passing through a single-mode fiber, the beam enters the core of the setup: the self-imaging cavity. A polarizing beam splitter (PBS) and a quarter-wave plate ($\nicefrac{\lambda}{4}$) image reflections onto a CMOS camera for alignment. The cavity consists of two partially reflecting mirrors, $M_1$ and $M_2$, and two bi-convex lenses, $L_2$ and $L_3$, arranged in a $4f$ configuration. The beam is focused by $L_1$ onto $M_1$, collimated by $L_2$, and passes through a sample in the cavity center. Lens $L_3$ and a post-cavity lens $L_4$ create a telescopic system with a magnification of 6.6x. This output is imaged via another $4f$-setup, optionally including a phase plate and an externally triggered camera (see \crefformat{figure}{Fig.~#2{A}#1#3}\cref{fig:setup2}).
Finally, a $70:30$ non-polarizing beamsplitter directs part of the signal to two photodiodes for monitoring.

In this work, we do not actively stabilize the cavity to a fixed length, but our setup allows for scanning across specific cavity resonances.  The driving voltage, which reflects the displacement, and the output signal are continuously monitored.
During the measurement, every $100\,\text{ms}$ we record a microscopy image with the CCD camera.

\section{Resolution in cavity-enhanced microscopy}

\begin{figure*}[t]
    \centering
    \includegraphics[width=\linewidth]{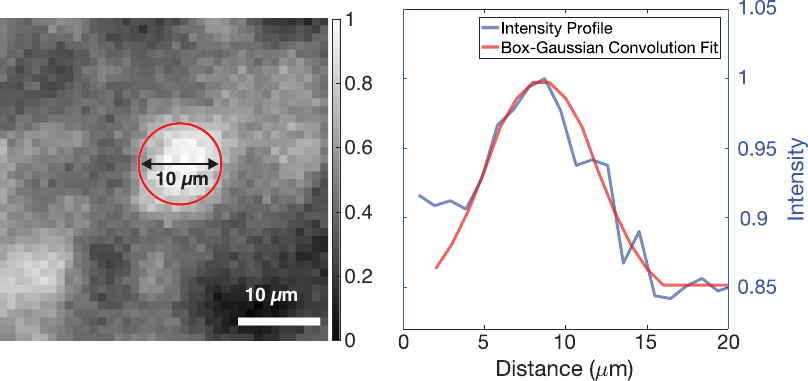}
    \caption{\textbf{Spot Size and Resolution Estimation} The left graphic shows the $10,\mu\text{m}$ hole when the cavity is on resonance. To assess the imaging capabilities of our setup, we aim to estimate the resolution. To do this, we fit a box function convolved with a Gaussian to the intensity profile of the hole, extracted from a single-pixel line cross-section. From this fit, we calculate the standard deviation, which serves as an estimate of the resolution. Theoretically, the lenses used can resolve features as small as approximately $2.8,\mu\text{m}$.}
    \label{fig:spot_res}
\end{figure*}

To estimate the resolution, we analyze the holes while the cavity is resonant. We are approaching this by modeling the hole as a box function. Due to the Gaussian blur induced by the used optics, the walls of the box profile are blurred as well. Consequently, the resolution can be determined by convolving the box function with a Gaussian. The resulting blur is dependent on $\sigma$, providing a reliable estimate of the resolution achieved.
Let $B(x)$ be a box function of width $2a$: \[ B(x) = \begin{cases} \frac{1}{2a} & \text{if } |x| \leq a \\ 0 & \text{otherwise} \end{cases} \] and $G(x)$ the Gaussian function: \[ G(x) = \frac{1}{\sqrt{2\pi}\sigma} e^{-\frac{x^2}{2\sigma^2}} \] with standard deviation $\sigma$. The convolution is then given by:
\[ f(x) = \int_{-\infty}^{\infty} B(t) G(x - t) \, dt = \frac{1}{4a} \left[ \operatorname{erf}\left(\frac{x + a}{\sqrt{2}\sigma}\right) - \operatorname{erf}\left(\frac{x - a}{\sqrt{2}\sigma}\right) \right] \] and $\sigma$ is found by fitting this model to the data of a hole intensity profile.
The hole size was measured by looking at the intensity plot in \ref{fig:spot_res}, which represents a cross section through the hole, yielding a standard deviation of $\sigma \approx 3 \mu m$. 

\section{Cavity-enhanced bright-field imaging of an optically thin sample - full multimode case}

Here we provide a linear-optics model for the bright-field signal of an optically thin, lossless sample in our 4f-cavity setup. We show that there is a gain in the contrast of the detected signal to local variations of the sample's phase shift if the imaging cavity is set to resonant transmission, as demonstrated in the main text for a $10\,$nm thin, punctured $\text{Si}_3\text{N}_4$ membrane. We also show that half of the contrast gain persists if the signal is averaged over a free spectral range of the imaging cavity, representing a scenario in which the cavity mirror positions are unstable over the course of the measurement.

\subsection{The 4f-cavity setup and model assumptions}

\begin{figure*}[t]
	\centering
	\includegraphics[width=\linewidth]{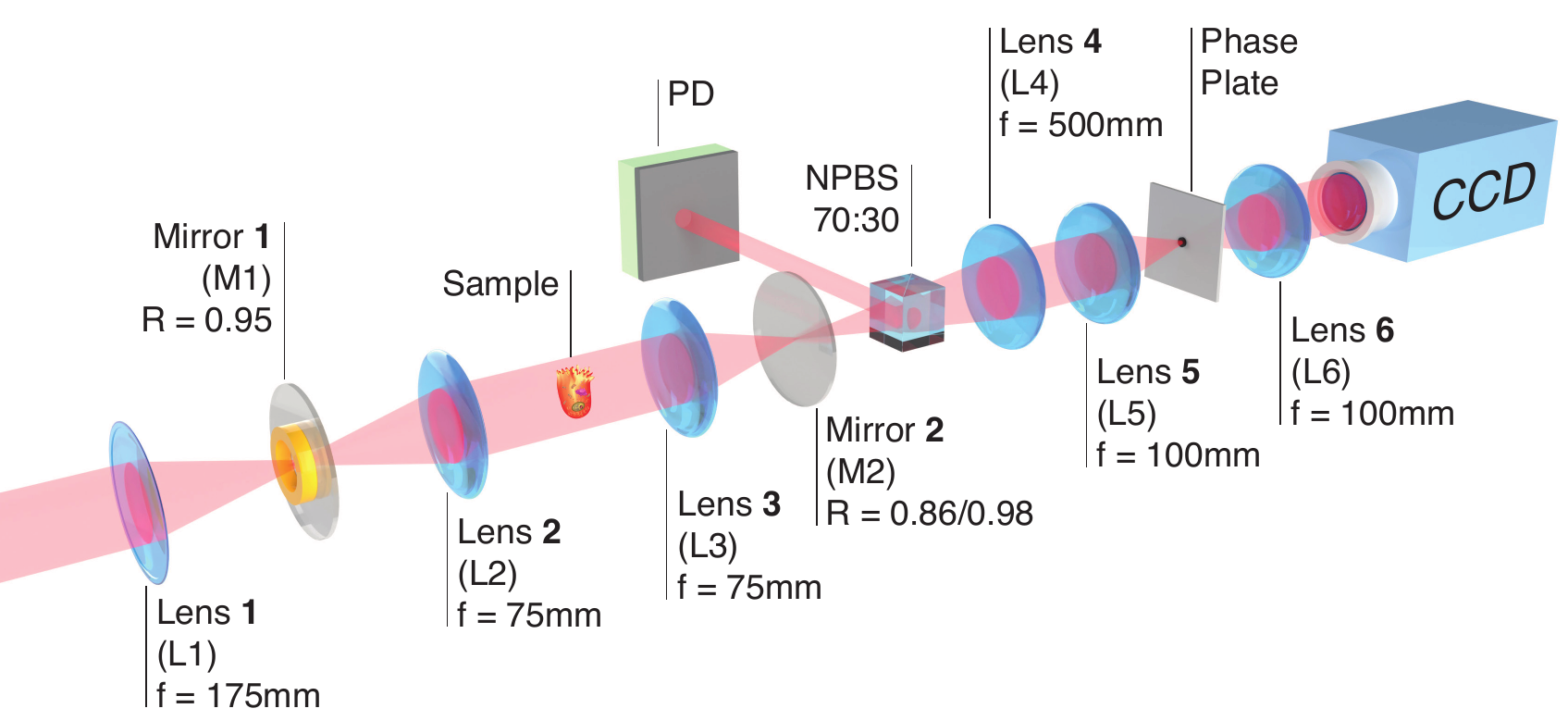}  
	\caption{ \textbf{Cavity setup} The self-imaging cavity is formed by two mirrors (M1 and M2) including lenses (L2 and L3) constituting a $4f$-setup. The cavity length is precisely controlled via a piezo ring mounted on M1. Light coupled into the resonator is focused on M1 to facilitate wide-field imaging in the sample plane, which is situated in the focal plane of lenses L2 and L3. After out-coupling, a non-polarizing beam splitter (NPBS) directs a faction of the light to a photodiode for output monitoring. Lens 4 provides a magnification of $6.6$, while an additional $4f$-setup enables various imaging techniques, including phase contrast. The resulting image of the sample is then detected by a CCD.}
	\label{fig:setup2}
\end{figure*}

The cavity setup is depicted in Fig.~\ref{fig:setup2}. It consists of two mirrors M1 and M2 with reflectivities $R_1$ and $R_2$ and two thin lenses of focal length $f$ in between, in a $4f$-arrangement. The position of mirror M1 can deviate by a few wavelengths $\lambda \ll f$ from its ideal position in the focal plane of the first lens. The sample, positioned in the central focus plane of both lenses, shall be described by a two-dimensional structure with thickness $d_s$ and real-valued refractive index $n_s (\vxy)$. We will be concerned with spatial variations above the diffraction limit. Here and throughout, $\vxy = (x,y)$ denotes the transverse coordinates on the any plane perpendicular to the optical axis, and $z$ denotes the optical axis coordinate. We make the following assumptions:

\begin{itemize}
    \item The light that probes the sample is a continuous-wave field (or sufficiently long pulse) of wavelength $\lambda$ in the paraxial regime, as described by a complex electric field amplitude $E(\vxy) e^{ikz}$ with wave number $k = 2\pi/\lambda$. We will omit the complex exponent and state all field amplitudes with respect to a reference plane at $z\equiv 0$ (say, the left focal plane of the first intracavity lens).
    \item For the reflection coefficients of the two mirrors M1 and M2, we choose the convention that the electric field interferes destructively with its reflection on the respective surfaces facing \textit{inside} the cavity, $r_{1,2} = -\sqrt{R_{1,2}}$. The reflection coefficients on the outside surfaces are thus of opposite, positive sign, and the transmission coefficients are $t_{1,2} = \sqrt{1-R_{1,2}}$. 
    \item The lenses are assumed to be ideally thin, and we neglect any resolution limit given by their aperture or other imperfections. Given the electric field profile $E(\vxy)$ illuminating the lens from one focal plane, the image on the opposite focal plane is then given by a Fourier transformation,
    \begin{equation}
    E_{2f} (\vxy) = \frac{-ik}{2\pi f} e^{2ikf} \widetilde{E} \left( \frac{k}{f}\vxy \right), \qquad \text{with} \quad \widetilde{E}(\vq) = \int \diff^2 r_{\perp} E(\vxy) e^{-i\vq\cdot\vxy}. \label{eq:2f_trafo}
    \end{equation}
    A strongly focused spot transforms into a wide collimated beam and vice versa.
    A sequence of two $2f$-transformations (as e.g.~experienced by the field reflected off a cavity mirror) results in an inversion of the field amplitude,
    \begin{equation}
    E_{4f} (\vxy) = -\left( \frac{k}{2\pi f}\right)^2 e^{4ikf} \widetilde{\widetilde{E}} \left( \frac{k}{f}\vxy \right) = -e^{4ikf} E(-\vxy) . \label{eq:4f_trafo}
    \end{equation}
    \item Given a paraxial beam and a free-standing sample structure sufficiently coarse compared to the light wavelength so that the imaging resolution is not limited by diffraction, we can describe the sample response by the position-dependent reflection and transmission coefficients of a dielectric slab with varying refractive index $n_s (\vxy)$,
        \begin{align}
        t(\vxy) &= \frac{4 n_s (\vxy) e^{i[n_s (\vxy)-1]kd_s (\vxy)}}{[n_s (\vxy)+1]^2 - [n_s (\vxy)-1]^2 e^{2in_s (\vxy)k d_s (\vxy) }}, \label{eq:t_sample} \\
        r(\vxy) &= \frac{ [n_s^2 (\vxy)-1] e^{-ik d_s (\vxy)} \left[e^{2in_s (\vxy)kd_s (\vxy)}-1\right]}{[n_s (\vxy)+1]^2 - [n_s (\vxy)-1]^2 e^{2in_s (\vxy)k d_s (\vxy) }}. \label{eq:r_sample}
    \end{align}
    Notice that $r(\vxy)$ is the same for reflection off both sides.
    \item The input field $E_{\rm in} (\vxy)$, defined as the amplitude illuminating the mirror M1 from the left, shall illuminate the relevant parts of the sample more or less homogeneously. In particular, we assume that the illumination is spatially symmetric, $E_{\rm in} (\vxy) = E_{\rm in} (-\vxy)$.
    \item We allow the left mirror position to deviate from the focal plane by $\delta z_1 \sim \lambda$, but we neglect the influence of the shift on the imaging of the field (since $\delta z_1 \ll f$). That is, the shift will change the (empty) cavity length to $kL = 4kf-\phi_1$ with $\phi_1 = k \delta z_1$ influencing the cavity resonance; however, the shift is still small enough to neglect any defocusing caused by it. Notice that the shift also implies that the sample is no longer in the cavity center, but rather displaced by $-\delta z_1/2$. 
\end{itemize}

\subsection{Derivation of the output field and detection signal}

Let $E_{\rm in} (\vxy)$ be the input field amplitude outside the left mirror M1, on the focal plane of the left intracavity lens. Including the phase shift $\phi_1 = k\delta z_1$ due to the mirror displacement by $\delta z_1$, and given the yet to be calculated backward-running wave from the sample that is reflected off M1 from the right, $E_{1 \leftarrow} (\vxy)$, the forward-running wave field on the right of M1 is then the sum of the transmitted and the reflected components,
\begin{equation}
    E_{1\rightarrow}(\vxy) = t_1 E_{\rm in} (\vxy) e^{i\phi_1} + r_1 E_{1 \leftarrow} (\vxy) . \label{eq:E_M1}
\end{equation}

The two running-wave fields on the right side of M1 are related to the running-wave fields on the left side of the sample plane via the $2f$-transform \eqref{eq:2f_trafo} and a phase shift by $-\phi_1$ due to the slightly shorter or longer distance $2f - \delta z_1$,
\begin{equation}
    E_{\rm L \rightarrow} (\vxy) = -\frac{ik}{2\pi f} e^{2ikf-i\phi_1} \widetilde{E}_{1\rightarrow} \left( \frac{k}{f}\vxy \right), \qquad E_{1 \leftarrow} (\vxy) = -\frac{ik}{2\pi f} e^{2ikf-i\phi_1} \widetilde{E}_{\rm L\leftarrow} \left( \frac{k}{f}\vxy \right).
    \label{eq:E_M1_S}
\end{equation}
Putting \eqref{eq:E_M1} and \eqref{eq:E_M1_S} together and invoking \eqref{eq:4f_trafo}, we have
\begin{equation}
    E_{\rm L \rightarrow} (\vxy) = E_0 (\vxy) - r_1 e^{4ikf-2i\phi_1}  E_{\rm L \leftarrow} (-\vxy), \qquad \text{with} \quad E_0 (\vxy) := -\frac{ikt_1}{2\pi f} e^{2ikf} \widetilde{E}_{\rm in} \left( \frac{k}{f}\vxy \right). \label{eq:EL_S}
\end{equation}
To the right of the sample plane, we can again relate the running-wave components to the ones left of M2 through a $2f$-transform. The difference is that there is no input field impinging on M2 from the right, so that $E_{2\leftarrow} (\vxy) = r_2 E_{2\rightarrow} (\vxy)$ and $E_{2\rightarrow} (\vxy) = (-ik/2\pi f) e^{2ikf}\widetilde{E}_{\rm R \rightarrow} (k\vxy/f)$. From this follows the relation between the waves right of the sample plane and the expression for the output field that we detect,
\begin{equation}
    E_{\rm R \leftarrow} (\vxy) = -r_2 e^{4ikf}  E_{\rm R \rightarrow} (-\vxy),\qquad E_{\rm out} (\vxy) = -\frac{ikt_2}{2\pi f} e^{2ikf} \widetilde{E}_{\rm R \rightarrow} \left( \frac{k}{f}\vxy \right).  \label{eq:E_M2}
\end{equation}
In matrix notation, we can combine \eqref{eq:EL_S} and \eqref{eq:E_M2} into 
\begin{equation}
    \begin{bmatrix} E_{\rm L \rightarrow} (\vxy) \\ E_{\rm R \leftarrow} (\vxy) \end{bmatrix} = \begin{bmatrix} E_{0} (\vxy) \\ 0 \end{bmatrix} - e^{4ikf} \begin{bmatrix} 0 & r_1 e^{-2i\phi_1} \\ r_2 & 0 \end{bmatrix} \begin{bmatrix} E_{\rm R \rightarrow} (-\vxy) \\ E_{\rm L \leftarrow} (-\vxy) \end{bmatrix} . \label{eq:E_4f_LR}
\end{equation}
The sample can be described by a transformation matrix mapping the incoming to the outgoing field components according to the coefficients \eqref{eq:t_sample} and \eqref{eq:r_sample},
\begin{equation}
  \begin{bmatrix} E_{\rm R \rightarrow} (\vxy) \\ E_{\rm L \leftarrow} (\vxy) \end{bmatrix}  = \mS (\vxy) \begin{bmatrix} E_{\rm L \rightarrow} (\vxy) \\ E_{\rm R \leftarrow} (\vxy) \end{bmatrix}, 
  \qquad \text{with} \quad \mS (\vxy) = 
  \begin{bmatrix} t(\vxy) & r(\vxy) \\ r(\vxy) & t(\vxy) \end{bmatrix} .
  \label{eq:S_trafo}
\end{equation}
Notice that the matrix is unitary since $|t|^2+|r|^2 = 1$ and $r^* t + rt^* = 0$, as one can easily check. 
For clarity and brevity of notation, we shall now drop the argument $\vxy$ and denote the coefficients and the matrix at this position by $t,r$, and $\mS$, whereas we denote by $\bar{t},\bar{r}$ and $\bar{\mS}$ the respective terms at the opposite position $-\vxy$.  Moreover, we subsume $\tilde{r}_1 \equiv r_1 e^{-2i\phi_1}$.
Plugging the sample transformation for $-\vxy$ into \eqref{eq:E_4f_LR} and iterating the equation with our condition of symmetric illumination, $E_0 (-\vxy) = E_0 (\vxy)$, we obtain
\begin{align}
    \begin{bmatrix} E_{\rm L \rightarrow} (\vxy) \\ E_{\rm R \leftarrow} (\vxy) \end{bmatrix} &= \begin{bmatrix} E_{0} (\vxy) \\ 0 \end{bmatrix} - e^{4ikf} \begin{bmatrix} 0 & \tilde{r}_1 \\ r_2 & 0 \end{bmatrix} \bar{\mS} \begin{bmatrix} E_{\rm L \rightarrow} (-\vxy) \\ E_{\rm R \leftarrow} (-\vxy) \end{bmatrix} \nonumber \\
    &= \left( \openone - e^{4ikf} \begin{bmatrix} 0 & \tilde{r}_1 \\ r_2 & 0 \end{bmatrix} \bar{\mS} \right) \begin{bmatrix} E_{0} (\vxy) \\ 0 \end{bmatrix} + e^{8ikf} \begin{bmatrix} 0 & \tilde{r}_1 \\ r_2 & 0 \end{bmatrix} \bar{\mS} \begin{bmatrix} 0 & \tilde{r}_1 \\ r_2 & 0 \end{bmatrix} \mS \begin{bmatrix} E_{\rm L \rightarrow} (\vxy) \\ E_{\rm R \leftarrow} (\vxy) \end{bmatrix} \nonumber \\
    &= E_0(\vxy) \begin{bmatrix} 1-e^{4ikf} \tilde{r}_1 \bar{r} \\ -e^{4ikf} r_2 \bar{t} \end{bmatrix} + e^{8ikf} \begin{bmatrix}
    \tilde{r}_1 (\tilde{r}_1\bar{r} r+r_2 \bar{t} t) & \tilde{r}_1 (\tilde{r}_1\bar{r} t+r_2 \bar{t} r) \\ r_2 (\tilde{r}_1\bar{t} r+r_2 \bar{r} t) &  r_2 (\tilde{r}_1\bar{t} t+r_2 \bar{r} r)
    \end{bmatrix} 
    \begin{bmatrix} E_{\rm L \rightarrow} (\vxy) \\ E_{\rm R \leftarrow} (\vxy) \end{bmatrix} .
\end{align}
The resulting linear equations can be solved straightforwardly,
\begin{align}
    E_{\rm L \rightarrow} &= E_0 \frac{1-e^{4ikf} [\tilde{r}_1\bar{r}+e^{4ikf}r_2 (r_2\bar{r}r+\tilde{r}_1\bar{t} t) + e^{8ikf}\tilde{r}_1 r_2^2 r(\bar{t}^2-\bar{r}^2)]}{1 - e^{8ikf}[(\tilde{r}_1^2+r_2^2)\bar{r}r + 2\tilde{r}_1 r_2 \bar{t}t] + e^{16ikf}\tilde{r}_1^2r_2^2 (\bar{t}^2-\bar{r}^2)(t^2-r^2)}, \nonumber \\
    E_{\rm R \leftarrow} &= E_0 \frac{e^{4ikf} r_2 [-\bar{t}+e^{4ikf} (\tilde{r}_1\bar{t}r+r_2\bar{r} t) + e^{8ikf}\tilde{r}_1 r_2 t(\bar{t}^2-\bar{r}^2)]}{1 - e^{8ikf}[(\tilde{r}_1^2+r_2^2)\bar{r}r + 2\tilde{r}_1 r_2 \bar{t}t] + e^{16ikf}\tilde{r}_1^2r_2^2 (\bar{t}^2-\bar{r}^2)(t^2-r^2)} . \label{eq:ELR_in}
\end{align}
where we have dropped the argument $\vxy$ of the field amplitudes, too. Another application of $\mS$ leaves us with
\begin{align}
    E_{\rm R \rightarrow} &= E_0 \frac{t-e^{4ikf} [ r_2\bar{t} r + \tilde{r}_1 \bar{r} t +e^{4ikf} \tilde{r}_1 r_2 \bar{t}(t^2-r^2) ]}{1 - e^{8ikf}[(\tilde{r}_1^2+r_2^2)\bar{r}r + 2\tilde{r}_1 r_2 \bar{t}t] + e^{16ikf}\tilde{r}_1^2r_2^2 (\bar{t}^2-\bar{r}^2)(t^2-r^2)}, \nonumber \\
    E_{\rm L \leftarrow} &= E_0 \frac{r-e^{4ikf} [r_2 \bar{t} t + \tilde{r}_1 \bar{r}r - e^{4ikf} r_2^2 \bar{r} (t^2-r^2) - e^{8ikf} r_2^2\tilde{r}_1 (\bar{t}^2-\bar{r}^2)(t^2-r^2) ]}{1 - e^{8ikf}[(\tilde{r}_1^2+r_2^2)\bar{r}r + 2\tilde{r}_1 r_2 \bar{t}t] + e^{16ikf}\tilde{r}_1^2r_2^2 (\bar{t}^2-\bar{r}^2)(t^2-r^2)} . \label{eq:ELR_out}
\end{align}
The output field in \eqref{eq:E_M2} that leaves the cavity on the right undergoes another transformation before it is detected: it passes a magnification lens L4 of focal length $f_4$ and, optionally, another $4f$-imaging system for further manipulation. 
For completeness, let us perform the $2f$-transformation corresponding to L4 and an unmanipulated 4f-transformation corresponding to the imaging system with $f_5$, which yields the magnified bright-field image of the sample,
\begin{align}
    E_{\rm bf} (\vxy) &= -e^{4ikf_5} E_4 (-\vxy) = \frac{ik}{2\pi f_4} e^{4ikf_5+2ikf_4} \int \diff^2 \vxy' E_{\rm out} (\vxy') e^{ik\vxy\cdot \vxy'/f_4} \nonumber \\
    &= \frac{k^2 t_2}{4\pi^2 f f_4} e^{4ikf_5+2ikf_4 + 2ikf} \int \diff^2 \vxy' \diff^2 \vxy'' E_{\rm R\rightarrow} (\vxy'') e^{ik\vxy\cdot \vxy'/f_4-ik\vxy'\cdot\vxy''/f} \nonumber \\
    &= \frac{t_2 f}{f_4} e^{4ikf_5+2ikf_4 + 2ikf} E_{\rm R \rightarrow} \left( \frac{f}{f_4} \vxy \right) . \label{eq:E_bf}
\end{align}
Since the magnification and the prefactors do not change the relevant phase signal of the sample, we will them implicitly and simply write $E_{\rm bf} = C E_{\rm R\rightarrow}$.

\subsection{Linear response and signal of a weak sample}

We get a clear picture about the cavity enhancement if we consider a weak and thin sample, i.e., we expand the sample coefficients of a thin slab to first order in $n_s kd_s$, 
\begin{equation} \label{eq:sampleParam}
    r(\vxy) \approx i\chi(\vxy),\quad t(\vxy) \approx 1+i\chi(\vxy),
    \qquad \text{with} \quad \chi(\vxy) = \frac{n_s^2(\vxy) - 1}{2} kd_s \ll 1.
\end{equation}
For our purposes, only the first, leading order in $\chi$ and $\bar{\chi} = \chi(-\vxy)$ is relevant, which allows us to neglect all second-order reflection terms ($r^2,\bar{r}^2,r\bar{r}$) in \eqref{eq:ELR_in} and \eqref{eq:ELR_out},
\begin{align}
    \begin{bmatrix} E_{\rm L \rightarrow} \\ E_{\rm R \leftarrow} \end{bmatrix} &\approx \frac{E_0}{(1-e^{8ikf}\tilde{r}_1 r_2 \bar{t}t)^2} \begin{bmatrix}
        1-e^{8ikf}\tilde{r}_1 r_2 \bar{t}t - e^{4ikf} \tilde{r}_1(\bar{r}+ e^{8ikf} r_2^2 r\bar{t}^2) \\ 
        e^{4ikf} r_2 [-\bar{t}(1-e^{8ikf}\tilde{r}_1 r_2 \bar{t}t)+e^{4ikf} (\tilde{r}_1\bar{t}r+r_2\bar{r} t) ]
    \end{bmatrix}, \label{eq:ELR_in_weak} \\
    \begin{bmatrix} E_{\rm R \rightarrow} \\ E_{\rm L \leftarrow} \end{bmatrix} &\approx \frac{E_0}{(1-e^{8ikf}\tilde{r}_1 r_2 \bar{t}t)^2} \begin{bmatrix}
        t(1-e^{8ikf}\tilde{r}_1 r_2 \bar{t}t) - e^{4ikf}(r_2 \bar{t} r + \tilde{r}_1 \bar{r} t) \\ 
        r - e^{4ikf}r_2\bar{t} t (1-e^{8ikf}\tilde{r}_1 r_2 \bar{t}t) + e^{8ikf}r_2^2 \bar{r} t^2
    \end{bmatrix}. \label{eq:ELR_out_weak}    
\end{align}
Recall that these are the fields evaluated at $+\vxy$ on the sample plane, but we obtain the fields at $-\vxy$ from the same formula if we exchange $r,t \leftrightarrow \bar{r},\bar{t}$.

In order to see how the cavity can enhance the bright-field contrast of the sample response, we expand the relevant output field \eqref{eq:E_bf} to first joint order in $\chi$ and $\bar{\chi}$, 
\begin{align}
    E_{\rm bf} = C E_{\rm R\rightarrow} &\approx \frac{E_0}{1-e^{8ikf}\tilde{r}_1 r_2} \left[ 1 + \frac{1-e^{4ikf}r_2}{1-e^{8ikf}\tilde{r}_1 r_2} i(\chi - e^{4ikf}\tilde{r}_1 \bar{\chi}) \right] \nonumber \\
    &= \frac{E_0}{1-e^{8ikf-2i\phi_1}\sqrt{R_1 R_2}} \left[ 1 + \frac{1+e^{4ikf}\sqrt{R_2}}{1-e^{8ikf-2i\phi_1}\sqrt{R_1 R_2}} i(\chi + e^{4ikf-2i\phi_1}\sqrt{R_1} \bar{\chi}) \right], \label{eq:E_bf_1storder}
\end{align}
where we have used our convention for $r_{1,2} = -\sqrt{R_{1,2}}$.
The outer prefactor describes the enhanced intra-cavity field strength, and it appears once more as a prefactor to the sample response. The prefactor achieves its maximum $(1-\sqrt{R_1R_2})^{-1}$ when the mirrors are perfectly positioned such that the cavity length $kL = 4kf-\phi_1 = \ell \pi $ with $\ell \in \mathbb{N}$. For the parameters in the manuscript, $R_1 = 0.95$ and $R_2 = 0.86$, this results in an amplification of about 10. The bright-field amplitude and intensity are then, to leading order, 
\begin{align}
    E_{\rm bf}^{\max} &= \frac{CE_0}{1-\sqrt{R_1 R_2}} \left[ 1 + i \frac{\chi +\sqrt{R_1R_2}\bar\chi +(-)^{\ell} (e^{i\phi_1}\sqrt{R_2}\chi + e^{-i\phi_1}\sqrt{R_1}\bar\chi)}{1-\sqrt{R_1 R_2}} \right], \\
    |E_{\rm bf}^{\max}|^2 &= \frac{|C E_0|^2}{(1-\sqrt{R_1 R_2})^2} \left[ 1 + 2(-)^{\ell} \frac{\sqrt{R_1}\bar\chi - \sqrt{R_2}\chi}{1-\sqrt{R_1R_2}} \sin\phi_1\right] = \frac{|C E_0|^2}{(1-\sqrt{R_1 R_2})^2} \left[ 1 + 2 \frac{\sqrt{R_1}\bar\chi - \sqrt{R_2}\chi}{1-\sqrt{R_1R_2}} \sin 4kf \right]. \label{eq:Ibf_max}
\end{align}
This shows us that, in order to observe the optimal cavity-enhanced contrast for a purely phase-shifting sample, the mirror position $\delta z_1$ and the focal length $f$ must be tailored precisely such that both $kL = 4kf-\phi_1 = \ell \pi$ and $\sin \phi_1 = (-)^\ell \sin 4kf = \pm 1$. This implies that both $4kf$ and $\phi_1$ must be \textit{odd} multiples of $\pi/2$, such that $\tilde{r}_1 = +\sqrt{R_1}$ and the sample displacement from the cavity center is an odd multiple of $\lambda/8$. 
In practice, it is unlikely that one achieves this precise balance since one typically does not even know the focal length of the lenses on the wavelength level and one thus cannot position the sample precisely enough in between the lenses either. On the other hand, it is also very unlikely that the configuration happens to be such that $4kf$ is a multiple of $\pi$ and the first order response vanishes. In each of our measurements, we can therefore assume that $4kf$ takes some fixed unknown value in between; adjusting the mirror position for maximum overall bright-field intensity then amounts to setting $\phi_1 = 4kf - \ell \pi$.

In an unstable cavity, the phase shift $\phi_1$ can be random as the cavity mirrors are drifting or fluctuating in position. We can take this into account by averaging the bright-field intensity uniformly over $\phi_1$ or, equivalently, over $kL$ at fixed $4kf$. For given $kL,\phi_1$, the intensity associated to \eqref{eq:E_bf_1storder} is
\begin{align}
    |E_{\rm bf}|^2 =& \frac{|CE_0|^2}{1+R_1R_2 - 2\sqrt{R_1 R_2}\cos 2kL} \left[ 1 - 2\text{Im}\frac{(1+e^{ikL + i\phi_1}\sqrt{R_2})(\chi + e^{ikL-i\phi_1}\sqrt{R_1} \bar{\chi})(1-e^{-2ikL}\sqrt{R_1 R_2})}{1+R_1R_2 - 2\sqrt{R_1 R_2}\cos 2kL}\right] \nonumber \\
    =& \frac{|CE_0|^2}{1+R_1R_2 - 2\sqrt{R_1 R_2}\cos 2kL} \nonumber \\
    &\times \left[ 1 - 2 \frac{ (\chi+\bar\chi)\sqrt{R_1 R_2}\sin 2kL + \sqrt{R_1}\sin(2kL-4kf)(\bar\chi + R_2 \chi) +\sqrt{R_2}(\chi + R_1\bar\chi)\sin 4kf }{1+R_1R_2 - 2\sqrt{R_1 R_2}\cos 2kL}\right]
\end{align}
Averaging this $\pi$-periodic function of $kL$ uniformly over its period results in
\begin{align}
   \overline{|E_{\rm bf}|^2} &= \frac{|CE_0|^2}{1-R_1R_2} \left[ 1 - 2\frac{(1+R_1R_2)\sqrt{R_2}(\chi + R_1\bar\chi)\sin 4kf - 2R_1\sqrt{R_2}(\bar\chi+R_2\chi)\sin 4kf}{(1-R_1 R_2)^2} \right] \nonumber \\
   &=\frac{|CE_0|^2}{1-R_1R_2} \left[ 1 -2 \sqrt{R_2}\frac{\chi - R_1 \bar\chi}{1-R_1 R_2} \sin 4kf \right] \label{eq:Ibf_avg}
\end{align}
In our experiment, we can quantify the image contrast by comparing the output intensities of a relevant sample pixel for which $\chi \neq \bar\chi \equiv \chi_0$ and a reference pixel with $\chi=\bar\chi = \chi_0$. In our case, $\chi_0$ is the phase response of the $\text{Si}_3\text{N}_4$ membrane, whereas $\chi$ can be zero if we pick one of the holes in the membrane as our sample area. Taking as the contrast the magnitude of the difference in pixel intensities divided by the sum, we get at fixed $4kf$, 
\begin{align}
    \mathcal{C}_{\max} &\approx \frac{\sqrt{R_2}}{1-\sqrt{R_1R_2}} |(\chi-\chi_0)\sin 4kf| \approx \frac{2\sqrt{R_2}}{T_1+T_2} |(\chi-\chi_0)\sin 4kf|, \\
    \mathcal{C}_{\rm avg} &\approx \frac{\sqrt{R_2}}{1-R_1R_2} |(\chi-\chi_0)\sin 4kf| \approx \frac{\sqrt{R_2}}{T_1+T_2} |(\chi-\chi_0)\sin 4kf|,
\end{align}
to lowest order for the ideal case \eqref{eq:Ibf_max} and for the averaged case \eqref{eq:Ibf_avg}, respectively. On the right, we have also expanded to lowest order in the mirror transmission, $T_{1,2} = 1-R_{1,2} \ll 1$, which reveals a mere factor-two difference between the ideal and the averaged contrast. 

\begin{figure*}[t]
	\centering
	\includegraphics[width=\linewidth]{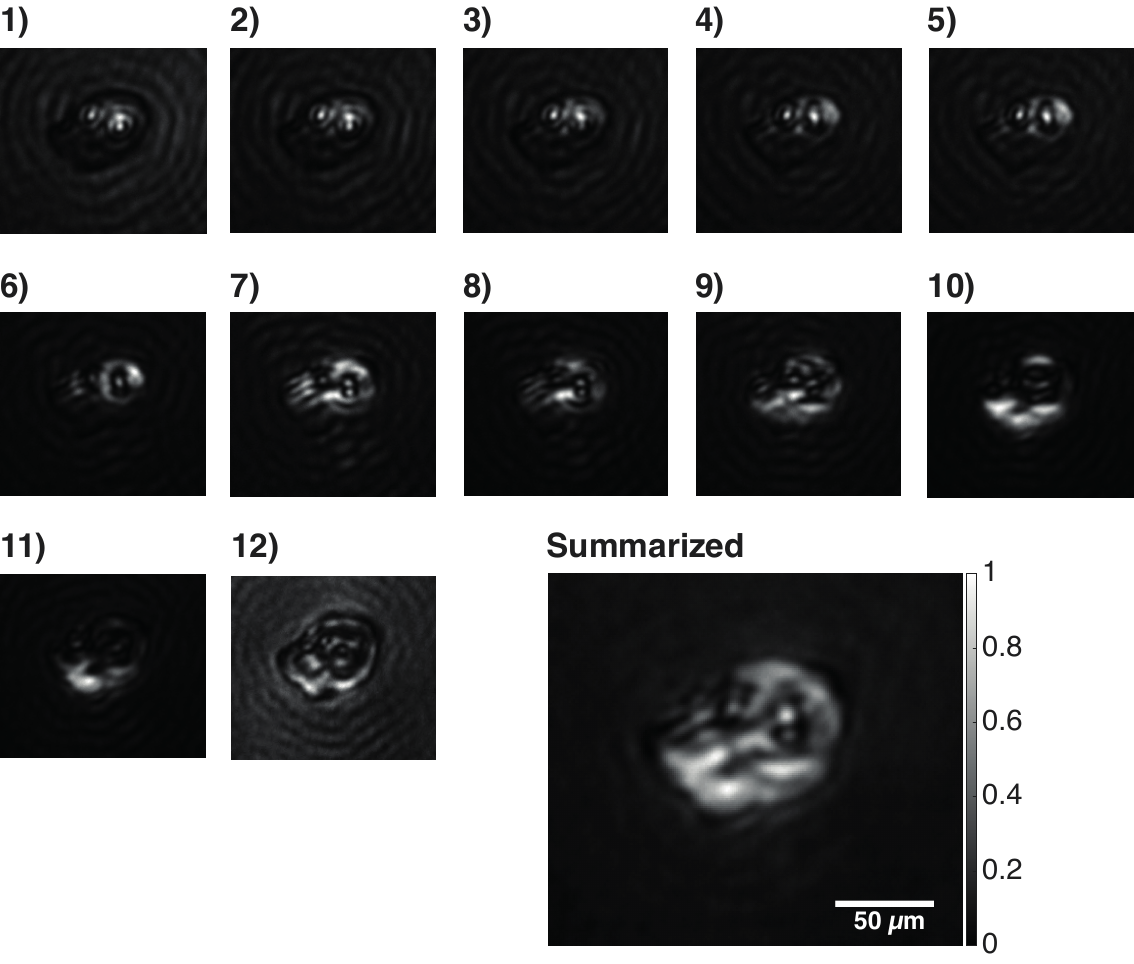}  
	\caption{\textbf{Cheek Cell in self imaging Cavity} Images 1 through 12 depict various sections of a cheek cell as the cavity length is adjusted. These images were captured while the probe light was off resonance. Different phase shifts result in distinct resonance points for the light passing through the cell, effectively enabling sensitive measurements of the local optical path length through the sample. When these images are combined, they create a dark-field representation of the cell. Refining this technique could provide a method for measuring sample thickness and examining specific regions in detail.}
	\label{fig:CC_cut}
\end{figure*}

\end{document}